# Experimental Verification of the Acoustic Geometric Phase

Bingyi Liu[1,2,†], Zhiling Zhou[3,†], Yongtian Wang[1], Thomas Zentgraf[2], Yong Li[3,‡] and Lingling Huang[1,§]

[1]*School of Optics and Photonics, Beijing Engineering Research Center of Mixed Reality and Advanced Display, Beijing Institute of Technology, Beijing, 100081, China*
[2]*Department of Physics, Paderborn University, Warburger Str. 100, 33098, Paderborn, Germany*
[3]*Institute of Acoustics, School of Physics Science and Engineering, Tongji University, Shanghai, 200092, China*

**Abstract**

The optical geometric phase encoded by the in-plane spatial orientation of microstructures has promoted the rapid development of numerous new-type optical meta-devices. However, pushing the concept of the geometric phase toward the acoustic community still faces challenges. In this work, we take advantage of two acoustic nonlocal metagratings that could support the direct conversion between plane wave and designated vortex mode, of which the orbital angular momentum conversion process plays a vital role in obtaining the acoustic geometric phase. We obtain the acoustic geometric phases of different orders by merely varying the orientation angle of one of the acoustic nonlocal metagratings. Intriguingly, according to our developed theory, we reveal that the reflective acoustic geometric phase, which is twice of the transmissive one, can be readily realized by transferring the transmitted configuration to a reflected one. Both the theoretical model and experimental measurements successfully verify the announced transmissive and reflective acoustic geometric phases. Moreover, the characteristics of reconfigurability and continuous phase modulation that covers the $2\pi$ range shown by the acoustic geometric phases provide us with new possibilities in advanced acoustic wavefront control.

---

[†] B. Liu and Z. Zhou contributed equally to this work.
[‡] Email: yongli@tongji.edu.cn
[§] Email: huanglingling@bit.edu.cn

Acoustic metasurfaces have been hailed as a revolutionary concept in renewing the comprehensive wave control with compact artificial scatters of high design flexibility.[1] Based on meticulously designed structures, acoustic metasurfaces can manipulate the phase, amplitude and velocity field of incident acoustic waves, which intrigues numerous fascinating applications in acoustic field engineering, such as beam generation,[2-6] holograms,[7-11] and noise reduction.[12-15] However, a considerable obstacle in the quest to achieve continuous phase modulation of the transmitted or reflected acoustic waves via metasurface structure design is the strong dependence of the transmission or reflection coefficient on the geometric parameters of structures, which generally requires a complex look-up table for suitable pixels in the meta-device design. Moreover, the phase modulation usually couples with the amplitude which inevitably deteriorates the performance of meta-devices.

In addition to the acoustic field engineering based on the local resonance phase shift achieved with the building pixel of metasurfaces, recently reported work on the acoustic geometric phase obtained via orbital angular momentum (OAM) transfer opens up another avenue in acoustic wave control by simply varying the orientation angle of the acoustic vortex elements, namely, the acoustic geometric-phase meta-array (AGPM),[16] which can be regarded as an acoustic version of the optical geometric-phase metasurface.[17-20] The acoustic geometric phase is well defined by the expression $\exp[i(l^{in} - l^{out})\varphi_{l^\xi}]$, where $l^{in}$ and $l^{out}$ refer to the topological charges (TCs) of the input and output acoustic vortex, respectively, and $\varphi_{l^\xi}$ is the rotational angle of the acoustic element with TC $l^\xi$, i.e., the acoustic vortex metasurface that is utilized for the OAM transfer process. For the simplicity of the following analysis, we define the value $l^{in} - l^{out}$ as the order of the acoustic geometric phase, and this term substantially describes the relation between the rotational angle and geometric phase. However, for the *n*-th order ($n \geq 2$) acoustic geometric phase, generally, at least three acoustic gradient type vortex metasurfaces are required (each of them accordingly accounts for vortex generation, conversion and detection), which makes the whole device bulky and decreases the transmission efficiency of the AGPM. In addition, the design of acoustic vortex metasurfaces usually involves complicated structures, e.g., Helmholtz resonators,[21-23] coiling channels,[24,25] and pentamode metamaterial,[26] where imperfect impedance matching of the building units together introduces additional reverberation and distorts the vortex conversion processes, especially the AGPM relying on higher-order acoustic geometric phases. Therefore, selecting an appropriate design strategy for the direct and high-efficiency vortex conversion between the plane wave and an arbitrary vortex would be a feasible solution for accurate acoustic geometric phase modulation of higher order, and also makes the whole system simple and compact.

Acoustic nonlocal metagratings show their nontrivial advantage in realizing highly efficient diffraction[27,28] and the generation of vortices with arbitrary TC,[29] where a simple and compact structure profile endows the device with robust performance by only numerically optimizing a few geometric parameters. Inspired by this, here we merely utilize two acoustic nonlocal metagratings and successfully realize 1st- to 3rd-order transmissive acoustic geometric-phase modulations; notably, this configuration can be directly applied to the reflection scenarios, where continuous reflective acoustic geometric phase modulation is also available according to our following theoretical analysis. Our work, for the first time, experimentally demonstrates a simple and efficient means to achieve flexible acoustic geometric-phase modulation, which opens up a new avenue in reconfigurable acoustic wave control.

Fig. 1 (a, b) show the typical transmissive and reflective acoustic geometric-phase pixels where continuous phase modulation can be obtained by varying the orientation angle of either the input (i.e., $\theta_i$, see Fig. 1(c)) or the output (i.e., $\theta_o$, see Fig. 1(d)) acoustic nonlocal metagrating. When we rotate the input and output acoustic nonlocal metagrating counterclockwise by an angle of $\varphi_i$ (see Fig. 1(c)) and $\varphi_o$ (see Fig. 1(d)), respectively, the overall geometric phase carried with the transmitted acoustic signal is $\exp[jl(\varphi_o - \varphi_i)]$, here $l$ is the TC of vortex generated by the acoustic nonlocal metagrating. In real application scenario, we generally fix the orientation angle of the input metagrating and only rotate the output metagrating by an angle of $\varphi_o = \varphi$, see Fig. 1(a, b). For the simplicity of the analysis, we assume that $\theta_i = \theta_o = \varphi_i = 0$ in this work, and the geometric phase carried by the transmitted acoustic wave is thereby $\exp(jl\varphi)$. Interestingly, when we simply seal the transmitted end with a rigid plug, the transmissive geometric-phase meta-atom given in Fig. 1(a) turns to a reflected one shown in Fig. 1(b), and the geometric phase carried by the reflected acoustic wave is twice of the transmissive geometric phase, i.e., $\exp(2jl\varphi)$. More details about the derivation of the transmissive and reflective acoustic geometric phase, which is obtained with two cascaded acoustic nonlocal metagratings, can be found in supplementary material.

Notably, the acoustic nonlocal metagratings show a great advantage in generating perfect acoustic vortices of nearly 100% conversion efficiency. Different from the conventional solution of constructing the phase gradient metasurface with precise manipulation of the local phase shifts of the subunits, the acoustic nonlocal metagrating whose double sides are decorated with fanlike grooves is rigorously designed based on the coupled mode theory (CMT),[29,30] where the precise manipulation of the energy flow between a given input and a target output mode is possible by optimizing multiple geometric parameters, which includes the number, depth and span angle of different fanlike grooves, and also the length of the central hole that connects the two sides of the nonlocal metagrating. The underlying physics of the nonlocal metagrating is pretty straightforward: the two-sided metagrating structure can be separately designed, which corresponds to two coherent perfect absorbers (CPAs) that account for the perfect absorption of the input mode (plane wave) and output mode (vortex mode).

Fig. 2(a) shows the schematic of an acoustic nonlocal metagrating and its key geometric parameters required for the next step optimization process. Here, the radius of the input and output waveguide is denoted by $R_i$ and $R_o$, respectively. The radius and height of the central hole is denoted by $r_c$ and $d_c$, respectively. Generally, given an example nonlocal metagrating designed for $|0\rangle \rightarrow |m\rangle$ conversion process and considering the design of the CPA for absorbing the incident mode $|m\rangle$. Here, the output side of nonlocal metagrating given in Fig. 2(a) is assumed to contain $I = I_r \times I_\varphi$ fanlike grooves, $I_r$ and $I_\varphi$ are the number of the grooves along the radial and polar direction, respectively. The depth of $(i,j)$ th $(i = 1,\cdots,I_r; j = 1,\cdots,I_\varphi)$ fanlike groove is denoted by $d_{ij}$, and the wall thickness along the radial and polar direction is denoted by $w_r$ and $w_\varphi$, respectively. The angular and radial span of the groove is denoted by $\Delta\varphi = (2\pi/I_\varphi) - w_\varphi$ and $\Delta r = [(R_o - r_c)/I_r] - w_r$. Moreover, the nonlocal metagrating has a polar period $a_\varphi$ and polar period number $p = 2\pi/a_\varphi$, where $|m| = p$. In this work, the design of CPA for absorbing a given incident vortex mode $|m\rangle$ is based on the CMT, and the study area is divided into three parts (see Fig. S1): Part I (light pink area): The output waveguide supporting the propagation of target vortex eigenmode $|m\rangle$; Part II (light green area): The grooves area; Part III (light blue area): The central hole area that links the input waveguide and output waveguide. Therefore, based on the continuum condition over the waveguide-metagrating interface at $z = 0$, we can finally determine the appropriate depths of the fanlike grooves and the length of the central hole with an optimization algorithm, more details about the design of acoustic nonlocal metagrating can be found in supplementary material.

According to our previously reported work,[29] we select the optimized geometric parameters for three types of nonlocal metagratings that support the direct OAM conversion, i.e., $|0\rangle \to |1\rangle$, $|0\rangle \to |2\rangle$ and $|0\rangle \to |3\rangle$ conversion processes. In this work, the operating frequency is selected as 3430 Hz and the corresponding wavelength $\lambda_0$ is 10 cm. For the simplicity of design, the radius $R_i$ of the input waveguide and the number of grooves of input surface of the above three nonlocal metagratings is fixed as 23.9 mm and 1×3, respectively, and the angular period is fixed as $a_\varphi = 2\pi$, and the radius $r_c$ of central hole is fixed as 11.9 mm. In addition, the radial wall thickness and polar span angle of wall are constant of 2.4 mm and 4.5°. The radius $R_0$ of the output waveguide that supports the propagation of vortex mode $|1\rangle$, $|2\rangle$ and $|3\rangle$ is 39.8 mm, 55.7 mm and 79.6 mm, respectively. The nonlocal metagrating and waveguide are assumed to be sound-hard, and a perfect matched layer is applied at the two far-end sides of the pixel to eliminate the reflected field. The depths of the fanlike grooves on output side and input side of each nonlocal metagrating can be found in supplementary material. Based on the commercial COMSOL Multiphysics finite element method (FEM) solver, Fig. 2 (b~d) show the numerical results of three typical nonlocal metagratings designed for the efficient generation of vortex with TC of 1, 2 and 3, which show relatively good helical pressure field distributions that carry desired TCs. The red and blue arrows refer to the input acoustic plane wave and output vortex fields, respectively.

We first numerically investigate the geometric phase modulation achieved with the two cascaded nonlocal acoustic metagratings based on the pressure acoustic module of COMSOL Multiphysics. Fig. 3 (a~c) show the geometric phase modulation achieved with the nonlocal metagratings given in Fig. 2(b~d), and a plane acoustic wave propagating along the $+z$ direction illuminates the pixel. Here, each pixel consists of two identical nonlocal metagratings with central inversion symmetry rather than simple mirror symmetry. When varying the orientation angle $\varphi$ of the output nonlocal metagrating counterclockwise from 0 to $2\pi$, the geometric phases carried by the output field are $\exp(j\varphi)$, $\exp(2j\varphi)$ and $\exp(3j\varphi)$, respectively. To ensure the conversion efficiency of the two nonlocal metagratings, the distance D among the input and output surfaces is set as $2.5\lambda_0$. For a much smaller distance (e.g., $0.5\lambda_0$), the interaction between two nonlocal metagratings is not negligible, which could cause unwanted reverberation and thereby distorts the linearity between the rotation angle of the output metagrating and geometric phase modulation.

Similarly, by truncating and sealing the output waveguide, the transmissive pixels given in Fig. 3 turn to the reflection modes. Fig. 3 (d~f) give the corresponding reflected geometric phase proportional to two folds of the transmissive geometric phase, i.e., $\exp(2j\varphi)$, $\exp(4j\varphi)$ and $\exp(6j\varphi)$, which agrees well with the conclusion derived from the above theoretical analysis. Both the transmissive and reflective geometric phases given in Figs. 3 show fairly good linearity regarding the orientation angle, while the amplitude of the transmitted or reflected coefficient is uniform and close to 1. In addition, the good linearity of the geometric phase modulation can be understood for the following two reasons: First, the nonlocal metagrating is rigorously designed for high-efficiency conversion between the plane and vortex beam. Second, a cylindrical waveguide possesses the cut-off frequencies for vortices of different TCs; therefore, the output waveguide of thin radius naturally serves as a good filter to support pure plane wave output.

Next, we experimentally verify the acoustic geometric phase predicted by our theory. The nonlocal metagrating, input waveguide and output waveguide are fabricated with the 3D printing technique whose printing accuracy is 0.1 mm, and the constituent material is photosensitive resin. Here, we anchor the input waveguide whose length is 250 mm with the nonlocal metagrating, and the output waveguide whose length is 250 mm is movable, more details about the fabrication of the sample can be found in supplementary materials. Fig. 4 (a) shows the schematic of the experiment setup. For the measurement of the transmissive acoustic geometric phase illustrated in Fig. 1(a), an absorbing foam is placed at the output end of the system to absorb the transmitted acoustic wave. Moreover, for the reflective acoustic geometric phase illustrated in Fig. 1(b), the absorbing foam is replaced by a solid plug, which is fabricated by 3D printing technique to well fit the waveguide and seals the far end of the system. In our experiment, the operating frequency is selected as the same as that used in simulations, i.e., $f_0 = 3430$ Hz, the amplitude and phase of either transmitted or reflected acoustic signal is measured with commercial LAN-XI data-acquisition kit (Brüel & Kjær type 3160-A-042) and two microphones (Brüel & Kjær type 2670). The nonlocal metagrating sample is rotated counterclockwise along the propagation direction of incident wave by step of 5°. Fig. 4(b) is the photograph of one set of the fabricated sample for experimental measurements.

Fig. 4(c) shows the experimental results of transmissive acoustic geometric phase (red solid dot) and transmitted amplitude (blue solid dot), which is obtained through the rotation of the second acoustic nonlocal metagrating that supports the $|l\rangle \rightarrow |0\rangle$ ($l = 1,2,3$) conversion process. It is obvious that the 1st, 2nd and 3rd order transmissive acoustic geometric phases are proportional to $\exp(j\varphi)$, $\exp(2j\varphi)$ and $\exp(3j\varphi)$, respectively, and $\varphi$ is the rotational angle of second nonlocal metagrating. By replacing the absorbing foam with a rigid plug, the system switches from transmission mode to reflection mode, and Fig. 4(d) gives the phase (red solid dot) and amplitude (blue solid dot) of the reflected acoustic signals when rotating the second nonlocal metagrating by an angle of $\varphi$, and the 1st, 2nd and 3rd order reflective acoustic geometric phases are proportional to $\exp(2j\varphi)$, $\exp(4j\varphi)$ and $\exp(6j\varphi)$, respectively, which are exactly two times of the corresponding transmitted geometric phases. Here, the amplitude of the reflected wave is generally smaller than that of the transmitted wave. This can be understood as the energy loss associated with the vortex conversions of the real nonlocal metagratings, and the reflected wave passes through the nonlocal metagratings for 4 times, while the transmitted wave only passes through two nonlocal metagratings. Notably, the acoustic geometric-phase modulation obtained with the two cascaded acoustic nonlocal metagratings offers a robust and flexible means in realizing a continuous phase control of the transmitted or reflected acoustic signals, which possess great potential in constructing the reconfigurable and even programmable active/passive acoustic wavefront modulation devices, more related discussions can be found in the supplementary material.

In conclusion, we theoretically analyze the arbitrary order transmissive and reflective acoustic geometric phase obtained with two identical nonlocal metagratings, and both numerical simulations and experimental measurements successfully verify the *n*-th order transmissive and reflective acoustic geometric phases. Our work offers a simple and compact platform for reconfigurable phase modulation of acoustic waves with the intriguing concept of acoustic geometric phase, which shows the potential in constructing reconfigurable acoustic meta-devices to generate complex fields.

**ACKNOWLEDGEMENTS**

The kind help offered by Dr. Yuhong Na (O) in preparing the artistic figures of the paper is acknowledged. Thanks for the helpful discussions with Dr. Qunshuo Wei and Dr. Zhaoxian Su. This project was supported by National Nature Science Foundation of China (Grant Nos. 12104044, 92050117), Beijing Outstanding Young Scientist Program (Grant No. BJJWZYJH0120190007022), China Postdoctoral Science Foundation (Grant No. 2021M690410) and Sino-German (CSC-DAAD) Postdoc

Scholarship Program, 2020 (Grant No. 57531629).

# AUTHOR DECLARATIONS

## Conflict of Interest

The authors have no conflicts to disclose.

# DATA AVAILABILITY

The data that support the findings of this study are available from the corresponding author upon reasonable request.

# SUPPLEMENTARY MATERIALS

See supplementary material for detailed discussions on the derivation of acoustic geometric phase, coupled mode theory utilized in the acoustic nonlocal metagrating design, geometry parameters and fabrication details of the samples we used in experiment, and beam generation functionality realized with the geometric-phase meta-atoms given in this work.

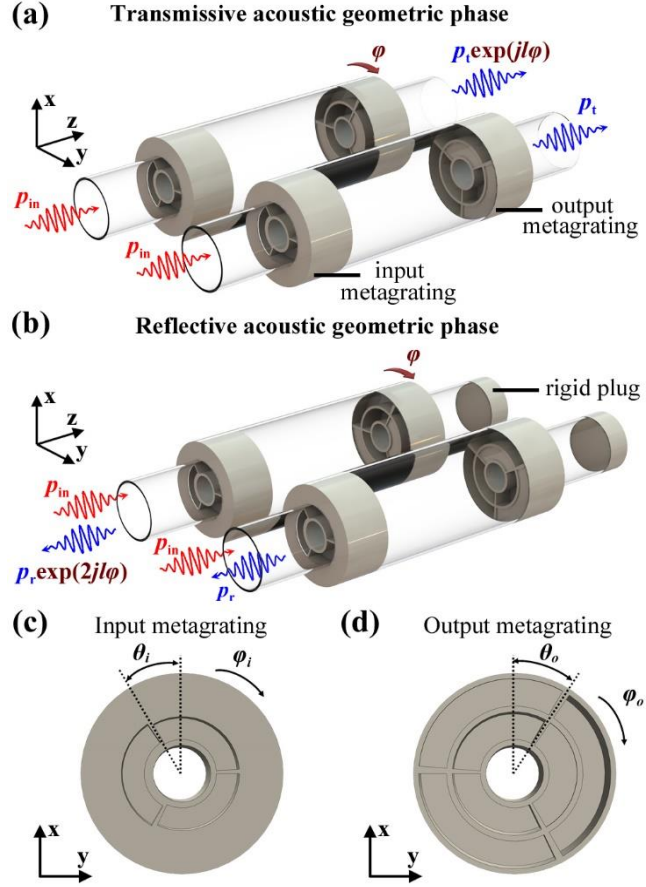

Fig. 1. Acoustic geometric phase based on two cascaded nonlocal acoustic metagratings that support the conversion between an acoustic plane wave and an acoustic vortex wave whose TC is *l*. (a) Transmissive and (b) reflective acoustic geometric phase modulation by varying the relative angle between the two nonlocal metagratings, i.e., fix the input metagrating and rotate the far end one by angle φ. The red pules refer to the input acoustic signals, the blue pulses given in (a) and (b) refer to the transmitted and reflected acoustic signals that carry the orientation angle-dependent geometric phase $\exp(j\varphi)$ and $\exp(2j\varphi)$, respectively. The top view of the (c) input and (d) output nonlocal metagrating, $\theta_i(\theta_o)$ and $\varphi_i(\varphi_o)$ refer to the initial orientation angle and rotational angle of the input (output) metagrating.

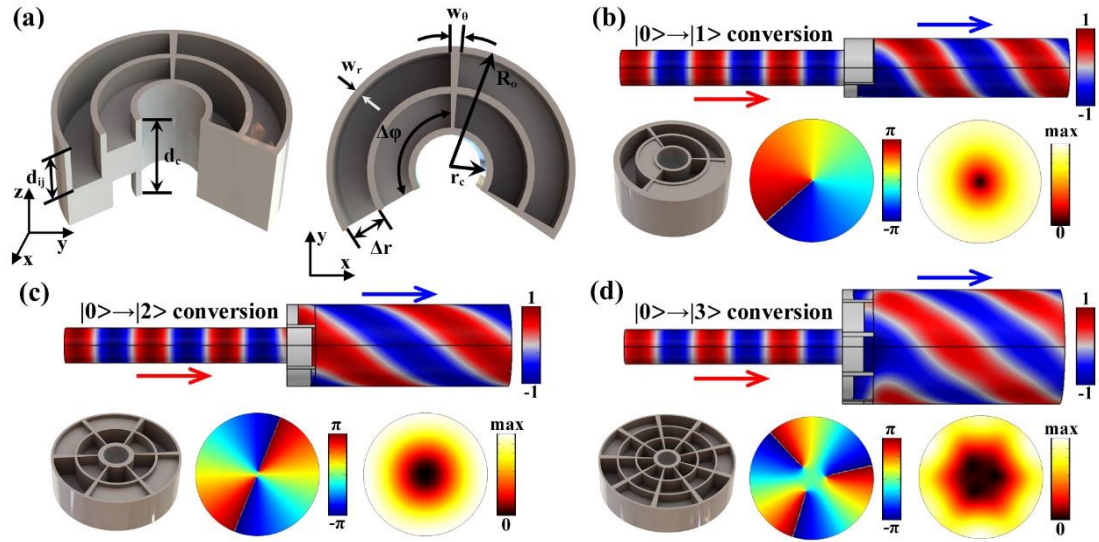

Fig. 2. Three acoustic nonlocal metagratings designed for high-efficiency vortex generation. (a) Schematic of acoustic nonlocal metagrating with key geometric parameters. Numerical simulation results for the (b) $|0\rangle \to |1\rangle$, (c) $|0\rangle \to |2\rangle$ and (d) $|0\rangle \to |3\rangle$ vortex conversion processes. For Fig.2. (b~d), the top panel shows the total acoustic pressure field when a plane acoustic wave is incident from an input waveguide to output waveguide through a nonlocal metagrating, the bottom left, middle and right panels show the schematics of nonlocal metagrating, phase and intensity distribution at the transmission port, respectively.

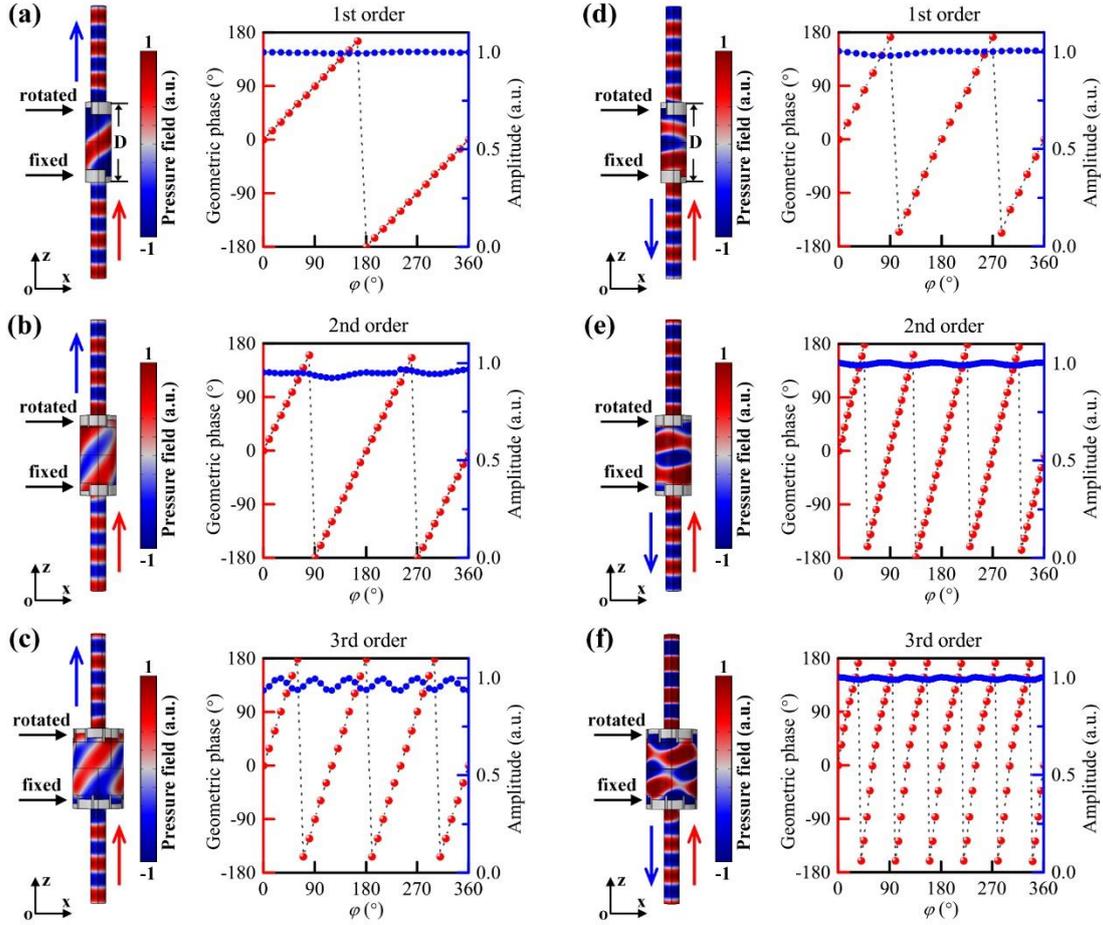

Fig. 3. Acoustic geometric phase obtained with two cascaded acoustic nonlocal metagratings. Transmissive acoustic geometric phase: (a)1st, (b) 2nd and (c) 3rd order transmissive geometric phases, which are proportional to $\exp(j\varphi)$, $\exp(2j\varphi)$ and $\exp(3j\varphi)$, respectively. Reflective acoustic geometric phase: (d)1st, (e) 2nd and (f) 3rd order reflective geometric phases, which are proportional to $\exp(2j\varphi)$, $\exp(4j\varphi)$ and $\exp(6j\varphi)$, respectively. The geometric phase modulation is obtained by counterclockwise rotating the top nonlocal metagrating by an angle $\varphi$. The red and blue arrows refer to the incident and transmitted waves, respectively. For each row, the left panel shows the numerical result of acoustic pressure field. The red and blue dots in the right panel refer to the normalized phase and amplitude of the transmitted coefficient, respectively.
14

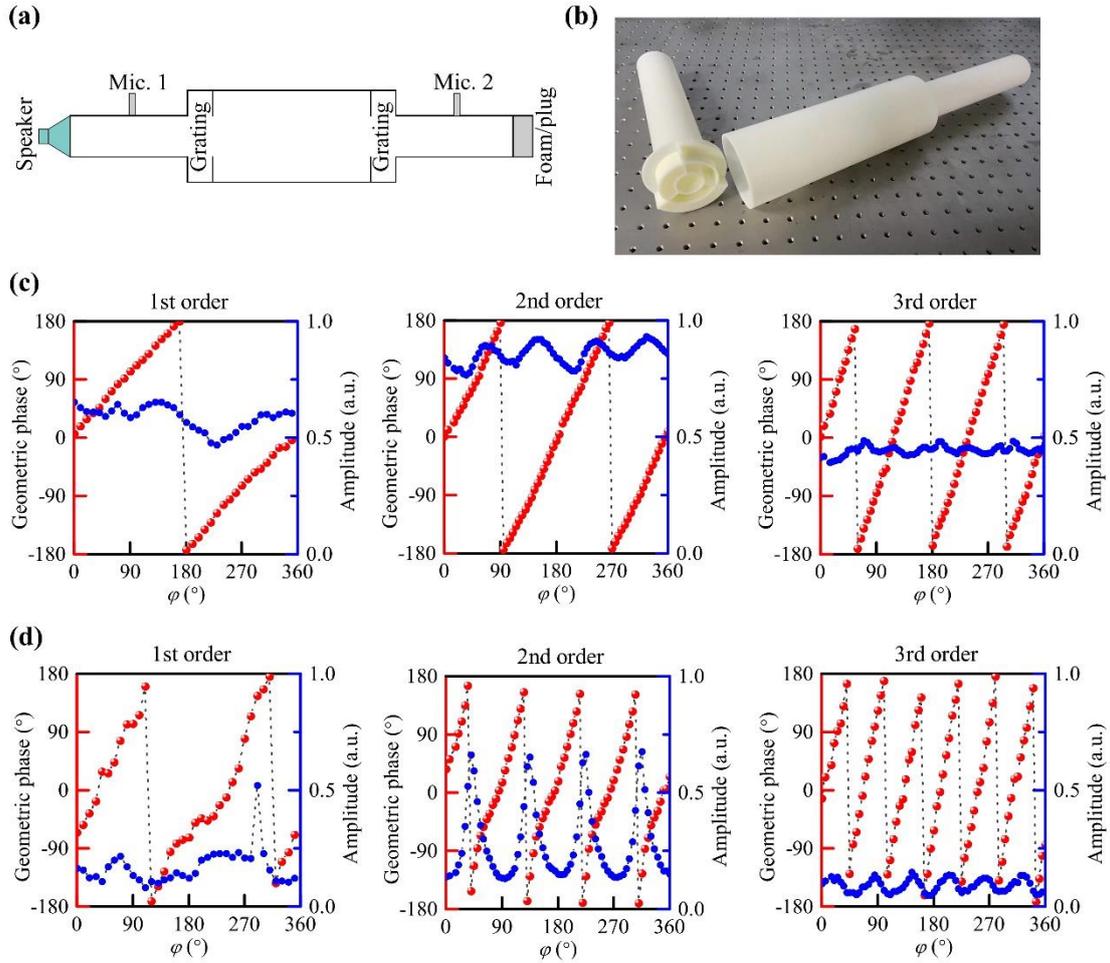

Fig. 4. Experimental verification of the acoustic geometric phase of different orders. (a) Schematic of the experimental setup. (b) Photograph of the fabricated sample. Experimental measured 1st to 3rd order (c) transmissive and (d) reflective acoustic geometric phase.